\documentclass[aps,pra,reprint,superscriptaddress,amsmath,amssymb,floatfix,nofootinbib,longbibliography]{revtex4-2}
\usepackage{graphicx}
\usepackage{amsmath}
\usepackage{hyperref} 
\usepackage{algorithm}
\usepackage{algpseudocode}

\newcommand{\BitAnd}{\,\&\,}
\newcommand{\BitXor}{\,\oplus\,}
\newcommand{\BitOr}{\,|\,}

\begin{document}

\title{\texorpdfstring{From Block Diagrams to Bloch Spheres: \\ Graphical Quantum Circuit Simulation in LabVIEW}{From Block Diagrams to Bloch Spheres: Graphical Quantum Circuit Simulation in LabVIEW}}

\author{Murtaza Vefadar}
\email{vafadar@gtu.edu.tr}
\affiliation{Department of Physics, Gebze Technical University, Kocaeli, Turkey}

\begin{abstract}
\vspace{0.5em}
As quantum computing transitions from theoretical physics to engineering applications, there is a growing need for accessible simulation tools that bridge the gap between abstract linear algebra and practical implementation. While text-based frameworks (like Qiskit or Cirq) are standard, they often present a steep learning curve for students and engineers accustomed to graphical system design. This paper introduces QuVI (Quantum Virtual Instrument), an open-source quantum circuit toolkit developed natively within the NI LabVIEW environment. Moving beyond initial proof-of-concept models, QuVI establishes a robust framework that leverages LabVIEW’s ``dataflow'' paradigm, in which wires represent data and nodes represent operations, to provide an intuitive, visual analog to standard quantum circuit notation while enabling the seamless integration of classical control structures like loops and conditionals. The toolkit's capabilities are demonstrated by constructing and visualizing fundamental quantum algorithms and verifying results against theoretical predictions. By translating ``Block Diagrams'' directly into quantum state evolutions (``Bloch Spheres''), QuVI offers researchers and educators a powerful platform for prototyping quantum logic without leaving the graphical engineering workspace.
\end{abstract}

\maketitle


\section{Introduction}
Quantum computing simulation serves as a crucial intermediary step between algorithm design and hardware execution \cite{nielsen, viamontes}. It allows researchers to verify logic, estimate resource requirements, and understand noise impacts in a controlled environment. Currently, the ecosystem is dominated by text-based frameworks such as Qiskit \cite{qiskit} and Cirq \cite{cirq}, which require a shift from the visual intuition of circuit diagrams to code.

Conversely, web-based graphical simulators like IBM Quantum Composer \cite{ibmcomposer} and Quirk \cite{quirk} provide intuitive, drag-and-drop interfaces for circuit construction. However, these tools generally lack robust integration with classical control logic. They are typically limited to static circuit execution or simple visualizations, making it difficult to implement dynamic, hybrid quantum-classical workflows involving conditional branching, iterative loops, or complex data processing.

LabVIEW (Laboratory Virtual Instrument Engineering Workbench) offers a unique alternative that bridges this gap. Its graphical programming language (G) is inherently parallel and isomorphic to quantum circuit diagrams: wires carry state information, and nodes perform operations. Unlike web-based simulators, LabVIEW is a full-featured programming environment. 

This work presents QuVI, a versatile quantum circuit toolkit for LabVIEW, which allows its users to wrap quantum circuits inside standard structures (such as For Loops, While Loops, and Case Structures), enabling the seamless construction of complex algorithms where classical logic controls quantum execution flow. A custom state-vector update engine is implemented to exploit CPU parallelism for both single-qubit controlled gates and multi-qubit operations, and the system is validated against standard quantum benchmarks. QuVI is publicly available via the VI Package Manager \footnote{\url{https://www.vipm.io/package/murtaza_vefadar_lib_quvi___quantum_circuit_toolkit/}}. 

\section{\texorpdfstring{Q\lowercase{u}VI Toolkit Architecture}{QuVI Toolkit Architecture}}
Unique to this implementation is the mapping of LabVIEW's graphical primitives to quantum circuit components. The simulation architecture relies on specific synchronization structures to maintain the global quantum state while permitting parallel execution of independent operations.

\subsection{State Management via Queues}
A critical challenge in graphical dataflow programming is managing large datasets without incurring memory copy overhead at every node. In QuVI, the visual wires connecting quantum gates do not carry the state vector data directly. Instead, a LabVIEW \textbf{Queue} of size 1 is utilized as a global storage buffer for the state vector and system parameters (e.g., qubit count). The wire transmits a reference to this queue, allowing gates to access and modify the shared state vector "in-place."

\subsection{Circuit Composition}
The simulation environment is constructed using a \textbf{Flat Sequence Structure}, which serves as the canvas for the quantum circuit.
\begin{enumerate}
    \item \textbf{Initialization:} The circuit begins with "QuVI Wire" subVIs, each initializing a qubit in the $|0\rangle$ state. These are stacked vertically to define the register size, following the convention that the topmost wire corresponds to $q_0$.
    \item \textbf{Wiring:} The horizontal output of each initialization VI is connected to the left border of the sequence structure, establishing the "quantum wires" that flow from left to right.
    \item \textbf{Execution Flow:} Inside the sequence structure, single-qubit gates placed on different wires execute independently via LabVIEW's native parallelism. Conversely, multi-qubit operations (e.g. controlled gates) introduce dependencies that require explicit synchronization mechanisms to ensure correct state evolution.
\end{enumerate}

\subsection{Synchronization with Notifiers}
\begin{figure}[htbp]
\centering
\includegraphics[width=0.9\columnwidth]{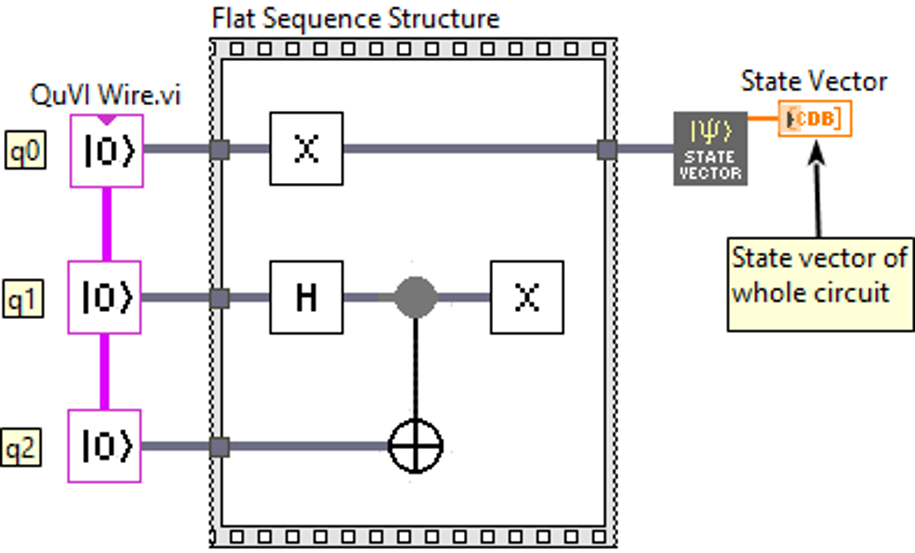}

\caption{QuVI architecture demonstrating execution flow. Top wire: independent operations execute in parallel via native dataflow. Bottom wires: entangling gates (e.g., CNOT) enforce synchronization using Notifiers to handle state dependencies between control and target qubits.}
\label{fig:arch_circuit}
\end{figure}

When entangling operations (controlled gates) occur, data dependencies arise between qubits (see Figure \ref{fig:arch_circuit}). To enforce this ordering, a ``Watch List'' mechanism is implemented. This list acts as a synchronization registry, stored in a LabVIEW Queue of size 1 containing an array of bits representing wire indices.

When a controlled gate is executed, the target gate first performs the necessary calculations on the global state vector. Once the update is complete, the Watch List bits are modified according to the control bitmask ("inclusion mask" in the next section). A LabVIEW Notifier is then employed to signal the waiting gates situated immediately after the control and anti-control nodes. Upon receiving this notification, these gates access the updated Watch List. Execution for a specific wire proceeds only if the corresponding bit in the array is observed to be ``off,'' allowing the gate to safely begin its own operations on the state vector.

\section{Methodology}
\vspace{0.5cm}
\subsection{Single Qubit Gate Update Algorithm}
To implement the simulation efficiently in a graphical environment, a perfectly parallelizable update scheme is utilized for arbitrary single-qubit gates (controlled or anti-controlled). This method iterates linearly through the state vector, a structure that maps directly to LabVIEW's Parallel For loops which automatically distribute iterations across processor cores.

\subsubsection{State Vector Update Rule}
The core of the QuVI engine avoids constructing the full Hilbert space operator $O = I \otimes \dots \otimes U \otimes \dots \otimes I$. Instead, it utilizes a per-element update strategy that is pleasingly parallel.
For a $2 \times 2$ gate matrix $U$, applied to target qubit $k$, the algorithm iterates through every index $i$ of the state vector independently. For each $i$, a ``partner'' index $j$ separated by a stride distance of $2^k$ (specifically, $j = i \oplus 2^k$) is identified.

\begin{figure*}[htbp]
\centering
\includegraphics[width=0.8\textwidth]{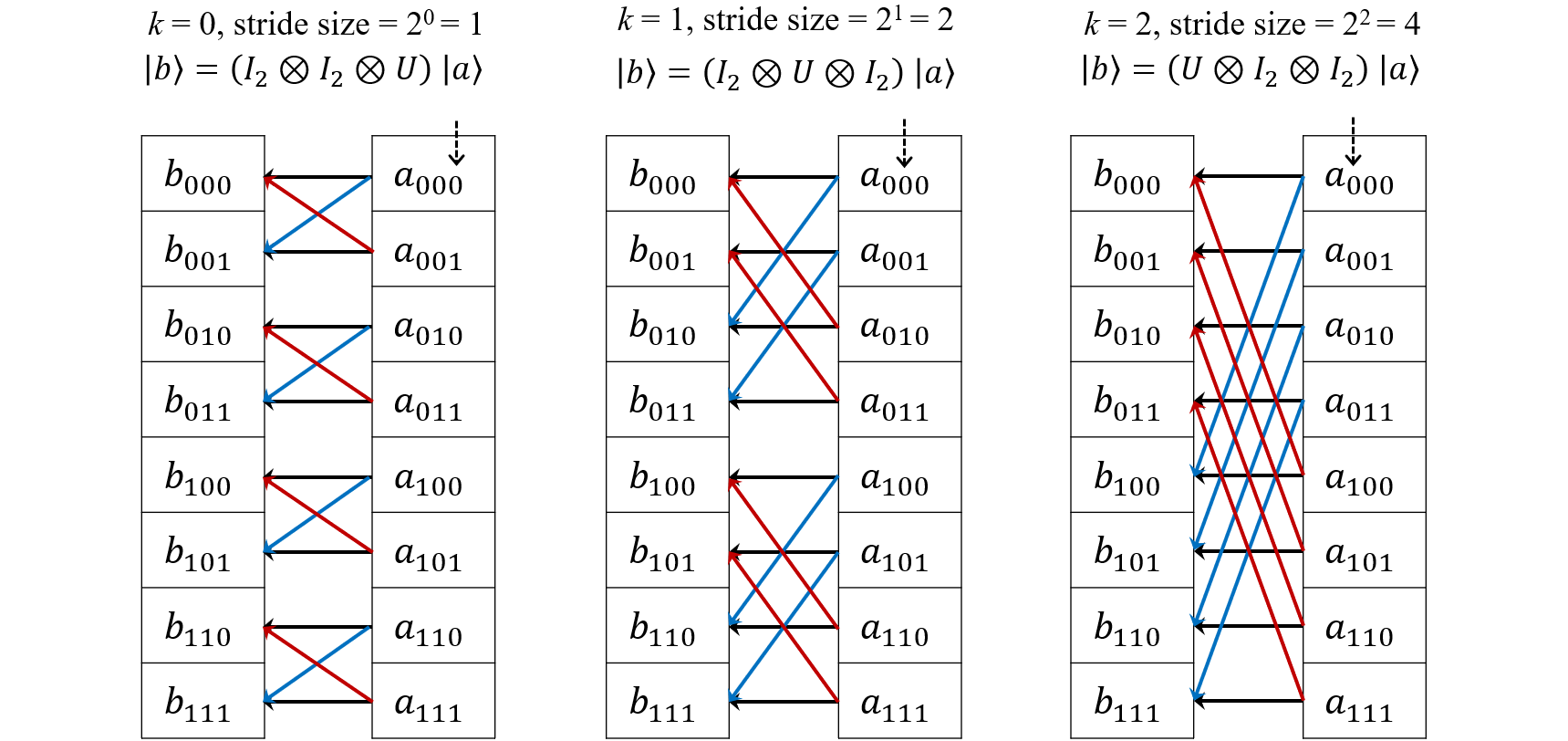}

\caption{Butterfly crossovers illustrating the stride logic for qubit $k$ (adapted from \cite{mcguffin}). The figure details the transformation formula of $|b\rangle$ from $|a\rangle$ for each case of a three-qubit system ($k=0, 1, 2$), where $I_2$ denotes the $2 \times 2$ Identity matrix. The arrow of the partner amplitude $a_{\text{partner}}$ is colored blue for $i_k=0$ and red for $i_k=1$ (dashed arrows mark $i_k$ indices for each case).}
\label{fig:butterfly}
\end{figure*}

The update logic calculates the new amplitude $b_i$ by mixing the input amplitudes $a_i$ and $a_j$ according to the gate matrix $U$. This relationship is illustrated by the butterfly structure in Figure \ref{fig:butterfly}. Because the computation of any output element $b_i$ relies only on the input state vector and does not depend on other output elements, the loop can be executed in parallel without race conditions.

Importantly, this approach minimizes memory overhead for the operator itself. Storing the full unitary matrix for an $n$-qubit system would require $O(4^n)$ memory  \cite{mcguffin}. In contrast, the kernel-based approach requires only $O(1)$ memory to store the gate parameters $U$, while maintaining the standard $O(N)$ memory requirement for the state vector itself (where $N=2^n$).

\subsubsection{Controlled Gate Logic}
To handle arbitrary control configurations, the update is generalized to a conditional element-wise rule using two bitmasks:
\begin{enumerate}
    \item \textbf{Inclusion Mask ($\mathcal{M}_{\text{inc}}$)}: A bitmask identifying which qubits act as controls. Bits are set to 1 at the positions of all control qubits.
    \item \textbf{Value Mask ($\mathcal{M}_{\text{val}}$)}: A bitmask specifying the required state for each control qubit. A bit is set to 1 for a standard control and 0 for an anti-control (conditioned on the qubit being $|0\rangle$).
\end{enumerate}

Let $\mathbf{a}, \mathbf{b} \in \mathbb{C}^{N}$ (where $N=2^n$) be the input and output state vectors, and let $U$ be a $2 \times 2$ unitary matrix acting on target qubit $k$. The update rule for every element $i \in \{0, \dots, N - 1\}$ is:

\begin{equation}
b_i =
\begin{cases}
    u_{00} a_i + u_{01} a_{\text{partner}} & \text{if } i \in \mathcal{C} \text{ and } i_k = 0 \\
    u_{10} a_{\text{partner}} + u_{11} a_i & \text{if } i \in \mathcal{C} \text{ and } i_k = 1 \\
    a_i & \text{if } i \notin \mathcal{C}
\end{cases}
\label{eq:update_rule}
\end{equation}

where the partner index is defined as $i \oplus 2^k$, and $\mathcal{C}$ is the set of indices satisfying the control condition $(i \BitAnd \mathcal{M}_{\text{inc}}) == \mathcal{M}_{\text{val}}$.

\subsubsection{Algorithm Implementation}
The implementation of this formulation is detailed in Algorithm \ref{alg:prll}. By calculating the partner index and control logic using bitwise operations, the branching logic remains local to each thread, minimizing synchronization overhead.

\begin{algorithm}[H]
\caption{Parallel Per-Element Single Qubit Gate Multiplication}
\label{alg:prll}
\begin{algorithmic}[1]
\Require 
\State $n$: Number of qubits
\State $\mathbf{a}$: Input State Vector (size $N = 2^n$)
\State $U$: $2\times2$ Unitary Matrix
\State $k$: Target Qubit Index
\State $\mathcal{M}_{\text{inc}}$: Inclusion Mask (Active Control Bits)
\State $\mathcal{M}_{\text{val}}$: Desired Value Mask (Control States)

\Ensure $\mathbf{b}$: Updated State Vector

\State $\text{targetStride} \gets 1 \ll k$
\State $N \gets 1 \ll n$ \Comment{Number of States}

\Statex \Comment{Main Loop: Perfectly Parallelizable in LabVIEW}
\For{$i \gets 0$ \textbf{to} $N - 1$} 
    
    \Statex \quad \Comment{Check Control Logic}
    \If{$(i \BitAnd \mathcal{M}_{\text{inc}}) == \mathcal{M}_{\text{val}}$}
        \State $\text{partner} \gets i \BitXor \text{targetStride}$
        
        \If{$(i \BitAnd \text{targetStride}) == 0$}
            \State \Comment{Case: Target bit is 0}
            \State $b[i] \gets (U_{00} \cdot a[i]) + (U_{01} \cdot a[\text{partner}])$
        \Else
            \State \Comment{Case: Target bit is 1}
            \State $b[i] \gets (U_{10} \cdot a[\text{partner}]) + (U_{11} \cdot a[i])$
        \EndIf
    \Else
        \State \Comment{Control condition failed. No change.}
        \State $b[i] \gets a[i]$
    \EndIf
\EndFor
\State \Return $\mathbf{b}$
\end{algorithmic}
\end{algorithm}

\subsection{Efficient SWAP Gate Implementation}
Simulating a SWAP gate typically requires exchanging states between two qubits $q_a$ and $q_b$. In a naive implementation, this might involve complex matrix permutations. However, another algorithm is utilized for the swap gate in a similar per-element parallel method, defining a direct mapping between input and output indices.

For a SWAP operation between qubits $q_a$ and $q_b$, the amplitude at index $i$ in the output vector corresponds to the amplitude at index $j$ in the input vector, where $j$ is obtained by swapping the bit values of $i$ at positions $a$ and $b$. This allows the construction of Algorithm \ref{alg:swap}, which remains perfectly parallelizable and supports arbitrary control masks.

\begin{algorithm}[H]
\caption{Parallel Swap Update}
\label{alg:swap}
\begin{algorithmic}[1]
\Require 
\State $n$: Number of qubits
\State $\mathbf{a}$: Input State Vector
\State $q_a, q_b$: Indices of qubits to SWAP
\State $\mathcal{M}_{\text{inc}}, \mathcal{M}_{\text{val}}$: Control Masks

\Ensure $\mathbf{b}$: Updated State Vector

\State $\text{maskA} \gets 1 \ll q_a$
\State $\text{maskB} \gets 1 \ll q_b$
\State $\text{swapMask} \gets \text{maskA} \BitOr \text{maskB}$
\State $N \gets 1 \ll n$

\Comment{Iterate over all states in parallel}
\For{$i \gets 0$ \textbf{to} $N - 1$} 
    \If{$(i \BitAnd \mathcal{M}_{\text{inc}}) == \mathcal{M}_{\text{val}}$}
        \State $\text{bitA} \gets (i \BitAnd \text{maskA}) \ne 0$
        \State $\text{bitB} \gets (i \BitAnd \text{maskB}) \ne 0$
        
        \If{$\text{bitA} \BitXor \text{bitB}$}
            \State $\text{source} \gets i \BitXor \text{swapMask}$
            \State $b[i] \gets a[\text{source}]$
        \Else
            \State $b[i] \gets a[i]$
        \EndIf
    \Else
        \State $b[i] \gets a[i]$
    \EndIf
\EndFor
\State \Return $\mathbf{b}$
\end{algorithmic}
\end{algorithm}

\section{Validation and Application Examples}
To demonstrate the capabilities of QuVI, we present two examples that highlight different aspects of the toolkit: classical feed-forward control in quantum teleportation, and iterative algorithmic structures in Grover's search.

\subsection{Quantum Teleportation}
The quantum teleportation protocol \cite{bennett} was constructed to transfer an arbitrary qubit state $|\psi\rangle = \alpha|0\rangle + \beta|1\rangle$ between two parties (Alice and Bob) using a Bell pair and classical communication.

\begin{figure*}[htbp]
\centering
\includegraphics[width=0.9\textwidth]{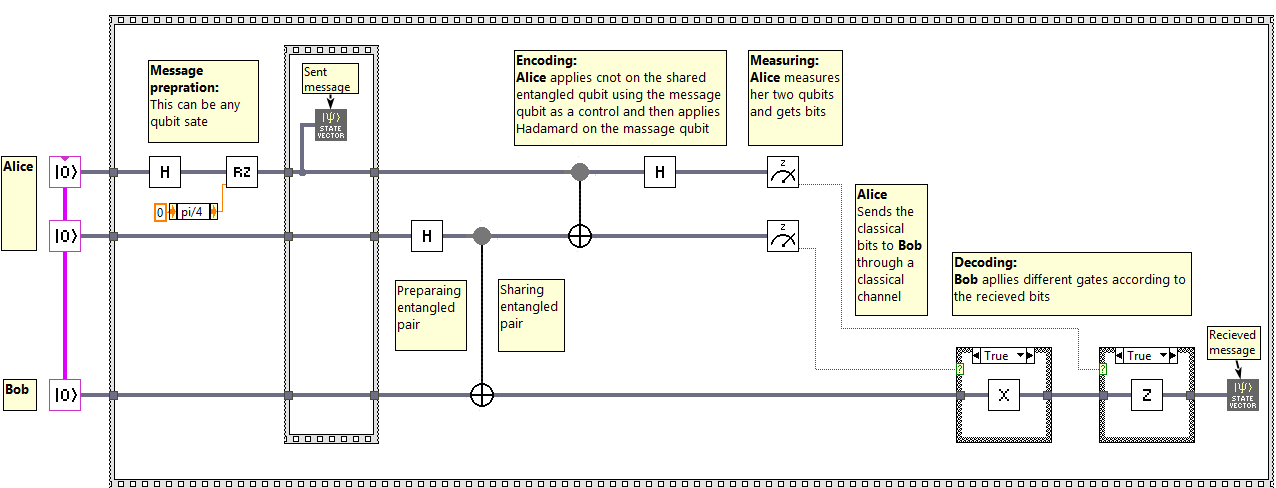}

\caption{Quantum Teleportation implemented in QuVI. The green dashed wires represent the classical channel, carrying measurement results from Alice's qubits to control Case Structures that apply corrections ($X$ and $Z$ gates) to Bob's qubit.}
\label{fig:teleportation}
\vspace{1cm}
\end{figure*}

Figure \ref{fig:teleportation} illustrates the implementation in QuVI. This example explicitly showcases the toolkit's "hybrid" nature. The measurement gates on Alice's side (top two wires) output boolean values which are transmitted via standard LabVIEW data wires (the green dashed lines acting as the "classical channel") to Bob's side. These boolean signals drive standard LabVIEW \textbf{Case Structures}, which conditionally apply the $X$ and $Z$ corrections to Bob's qubit. This effectively simulates the feed-forward logic required for teleportation without needing separate classical and quantum simulation environments.
\vspace{-7mm}

\subsection{Grover's Search Algorithm}
A 4-qubit Grover's search algorithm \cite{grover} was implemented to locate a marked state within a search space of $N=16$ items. This example highlights the use of LabVIEW's iterative structures to simplify circuit design.

\begin{figure*}[htbp]
\centering
\includegraphics[width=0.9\textwidth]{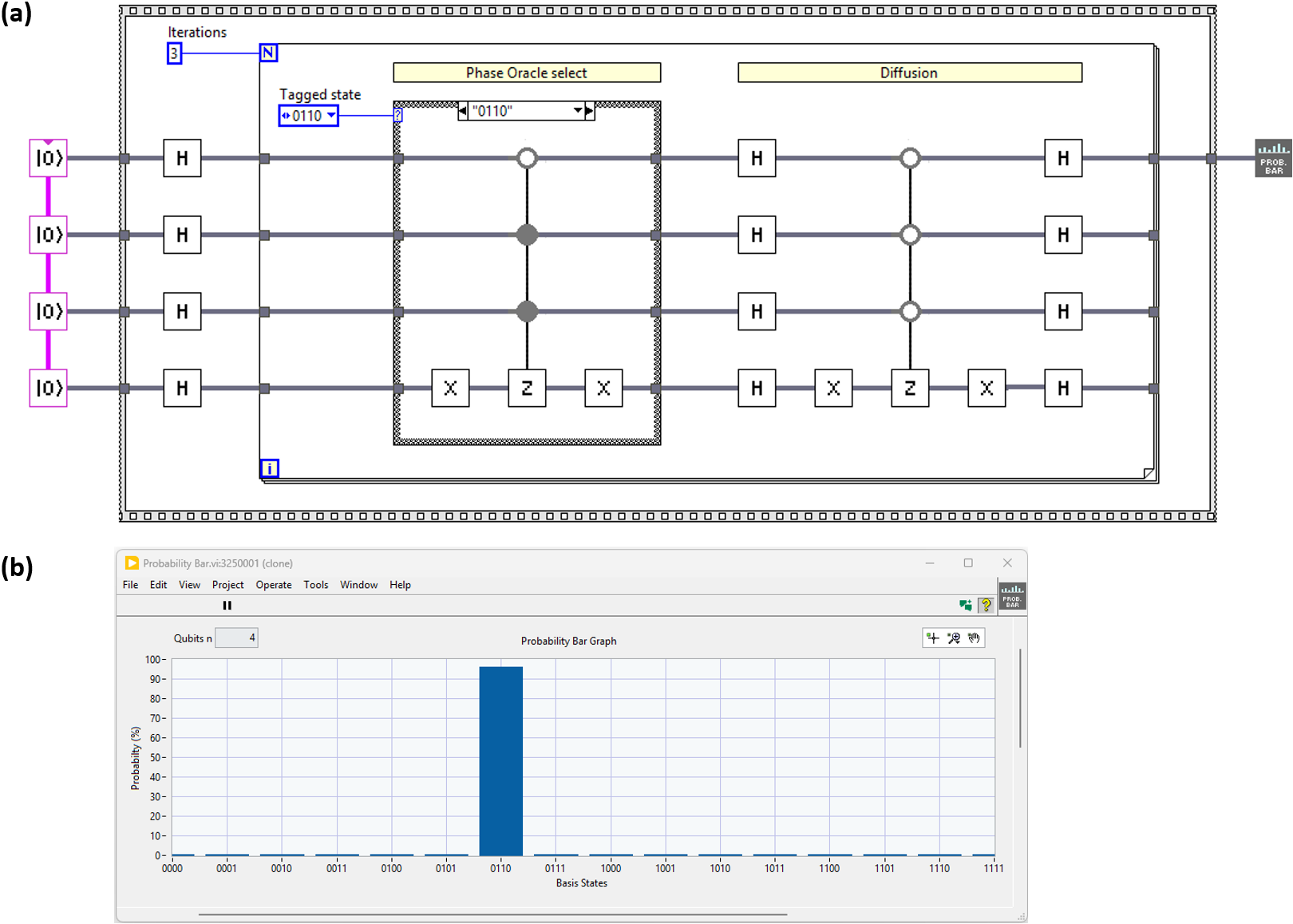}

\caption{4-Qubit Grover Search in QuVI. (a) LabVIEW Block Diagram utilizing a \textbf{For Loop} for algorithm iterations and a \textbf{Case Structure} for dynamic Oracle selection. The tagged state is set to $|0110\rangle$. (b) Simulation result showing the probability distribution of the final state, with a dominant peak ($>96\%$) at the index 0110.}
\label{fig:grover_diagram}
\vspace{1cm}
\end{figure*}

The core sequence of the algorithm, the Oracle ($O$) followed by the Diffusion operator ($D$), is placed inside a LabVIEW \textbf{For Loop} structure (see Figure \ref{fig:grover_diagram}). This allows the user to automate the amplitude amplification process, running the sequence $R \approx \frac{\pi}{4}\sqrt{N}$ times (here $R=3$) without manually duplicating the block diagram components. Additionally, a \textbf{Case Structure} is used to dynamically select the Phase Oracle based on a user input (here set to tagged state $|0110\rangle$), allowing a single VI to search for any target state.

The system was initialized in the equal superposition state $|s\rangle = H^{\otimes 4}|0\rangle$. After 3 iterations determined by the For Loop, the simulator correctly amplified the probability of measuring the dynamically selected target state ($|0110\rangle$) to $>96\%$, consistent with theoretical predictions.
\vspace{-5mm}

\section{Conclusion}

QuVI successfully demonstrates that a versatile quantum circuit simulator can be built natively in LabVIEW, offering robust integration between quantum operations and classical control logic. By implementing a pleasingly parallel per-element update engine within a synchronization architecture utilizing Queues and Notifiers, a modular system was achieved that adheres to the graphical dataflow paradigm while enabling complex, hybrid algorithm design. This tool provides a unique platform for engineering students and researchers to prototype and visualize quantum mechanics using familiar instrumentation metaphors.

Future enhancements will focus on extending the simulator beyond pure state vectors to support mixed states via the density matrix formalism. This development will enable the simulation of real, noisy quantum systems through the incorporation of decoherence channels. Furthermore, diagnostic tools for quantifying entanglement, such as von Neumann entropy and concurrence \cite{guhne}, and computing reduced density matrices via partial traces \cite{mcguffin} will be implemented.

\bibliographystyle{ieeetr}
\bibliography{refs}

\end{document}